\documentclass[sigconf,screen,natbib]{acmart}
\usepackage[utf8]{inputenc}
\usepackage{soul}
\usepackage{url}
\usepackage[utf8]{inputenc}
\usepackage{graphicx}
\usepackage{amsmath, amsfonts}
\usepackage{amsthm}
\usepackage{balance}
\usepackage{booktabs}
\usepackage{algorithmicx}
\usepackage[ruled]{algorithm}
\usepackage[noend]{algpseudocode}
\usepackage{multirow}

\copyrightyear{2022} 
\acmYear{2022} 
\setcopyright{acmcopyright}
\acmConference[CIKM '22] {Proceedings of the 31st ACM International Conference on Information and Knowledge Management}{October 17--21, 2022}{Atlanta, GA, USA.}
\acmBooktitle{Proceedings of the 31st ACM International Conference on Information and Knowledge Management (CIKM '22), October 17--21, 2022, Atlanta, GA, USA}
\acmPrice{15.00}
\acmISBN{978-1-4503-9236-5/22/10}
\acmDOI{10.1145/3511808.3557133}

\newcommand{\ie}{\textit{i.e.},\ }
\newcommand{\eg}{\textit{e.g.},\ }

\newcommand{\etc}{etc.\ }

\begin{document}

\title{Multi-Agent Reinforcement Learning for Network Load Balancing in Data Center}


\author{Zhiyuan Yao}
\orcid{0000-0002-7211-1506}
\affiliation{%
  \institution{\'Ecole Polytechnique \& Cisco Systems}
  \city{Paris}
  \state{}
  \country{France}
}
\email{zhiyuan.yao@polytechnique.edu}

\author{Zihan Ding}
\orcid{0000-0002-6294-888X}
\affiliation{%
  \institution{Princeton University}
  \city{Princeton}
  \state{New Jersey}
  \country{USA}
}
\email{zihand@princeton.edu}

\author{Thomas Clausen}
\orcid{0000-0002-7400-8887}
\affiliation{%
  \institution{\'Ecole Polytechnique}
  \city{Paris}
  \state{}
  \country{France}
}
\email{thomas.clausen@polytechnique.edu}

\begin{abstract}
This paper presents the network load balancing problem, a challenging real-world task for multi-agent reinforcement learning (MARL) methods.
Conventional heuristic solutions like Weighted-Cost Multi-Path (WCMP) and Local Shortest Queue (LSQ) are less flexible to the changing workload distributions and arrival rates, with a poor balance among multiple load balancers.
The cooperative network load balancing task is formulated as a Dec-POMDP problem, which naturally induces the MARL methods.
To bridge the reality gap for applying learning-based methods, all models are directly trained and evaluated on a real-world system from moderate- to large-scale setups.
Experimental evaluations show that the independent and ``selfish'' load balancing strategies are not necessarily the globally optimal ones, while the proposed MARL solution has a superior performance over different realistic settings.
Additionally, the potential difficulties of the application and deployment of MARL methods for network load balancing are analysed, which helps draw the attention of the learning and network communities to such challenges.
\end{abstract}



\maketitle

\section{Introduction}
\label{sec:intro}

\begin{figure}[t]
	\vskip 0.1in
	\begin{center}
		\centerline{\includegraphics[width=\columnwidth]{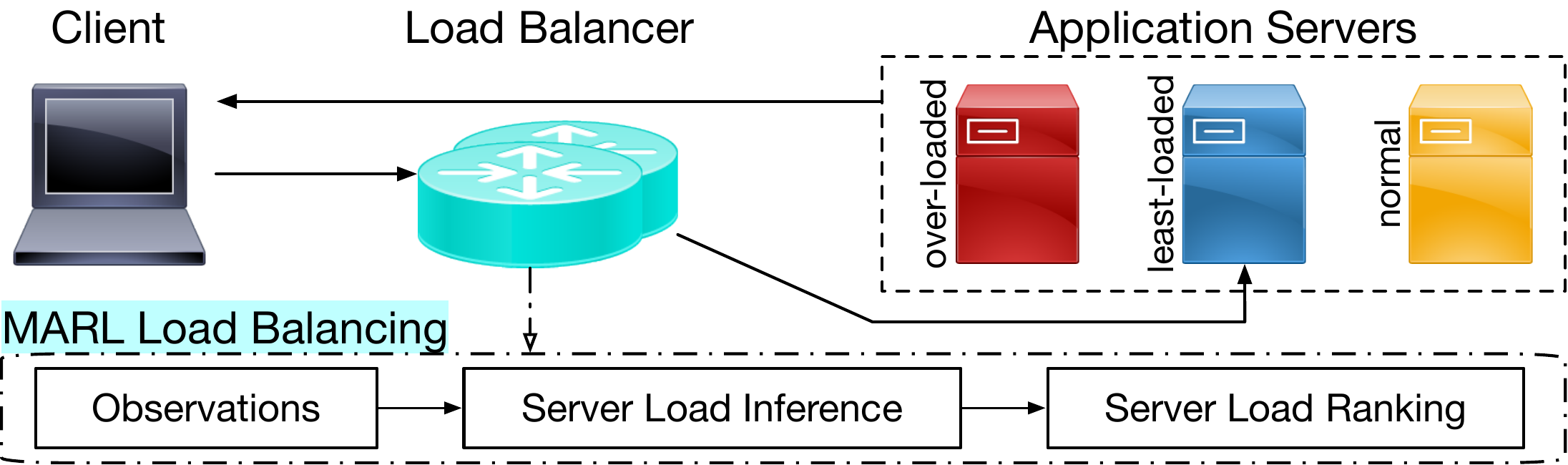}}
		\vskip -.1in
		\caption{Network load balancing in DC networks and the scope of study of this paper.}
		\label{fig:architecture}
	\end{center}
	\vskip -0.3in
\end{figure}

In data centers (DCs), network load balancers (LBs) play a significant role to distribute time-sensitive requests from clients across a cluster of application servers and provide scalable services~\cite{eisenbud2016maglev}.
The network topology of load balanced system in real-world DCs is depicted in Figure~\ref{fig:architecture}.
With the advancement of virtualisation technology and elastic data centers~\cite{dragoni2017microservices}, application servers can be instantiated on heterogeneous architectures~\cite{kumar2020fast} and have different processing capacities.
This requires network LBs to make fair workloads distribution decisions based on instant server load states to optimise resource utilisation, so that less application servers can be provisioned to provide better quality of service (QoS)--lower job completion time (JCT)--with reduced operational costs.
However, there are \textbf{2 challenges} for network LBs to make fair workloads distribution decisions in real-world systems.

\begin{figure}[t]
	\vskip 0.1in
	\begin{center}
		\centerline{\includegraphics[width=\columnwidth]{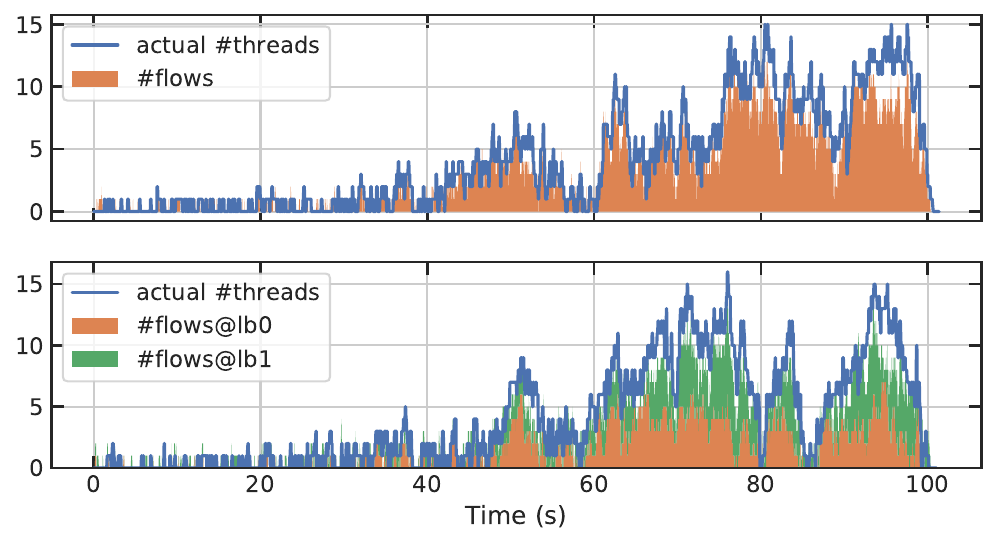}}
		\vskip -0.1in
		\caption{Comparison of the observed number of on-going flows (\#flow) between single- (top) and double-load-balancer (bottom) environment.}
		\label{fig:feature-1lb-vs-2lb-nf}
	\end{center}
	\vskip -0.3in
\end{figure}

\textbf{Network LBs have limited observations to make informed decisions.}
Operating at the Transport Layer, network LBs are agnostic to Application-Layer protocol and do not inspect the Application-Layer headers or payloads in network packets, in order to generalise to, and stay universal for all types of network applications~\cite{eisenbud2016maglev}.
However, this makes LBs also agnostic to the information of received requests and network flows (jobs)--\textit{e.g.}, expected JCT, computation intensity, required database--when making load balancing decisions, which can lead to overloaded servers dealing with multiple heavy network flows~\cite{twf2020}.
Besides, to avoid single-point-of-failure and improve system reliability, multiple LBs are deployed in modern DCs so that the service stays available when a LB fails.
To improve workload distribution fairness, heuristic LBs~\cite{twf2020} make informed load balancing decisions based on simple features (\eg the number of on-going connections) extracted from network packets.
However, the presence of multiple LBs makes them have only partial observations on the network traffic and workloads distributed among servers.
An example of partial observation of the number of on-going flows is depicted in Figure~\ref{fig:feature-1lb-vs-2lb-nf}.
In presence of $2$ LBs at the same time, the counted number of flows (\texttt{\#flow}) no longer reflect the actual number of busy threads (\texttt{\#threads}) on the server.
This indicates that more features--especially those that are not affected by partial observations (\eg latency-related features)--should be taken into account when making load balancing decisions.

\begin{figure}[t]
	\vskip 0.1in
	\begin{center}
		\centerline{\includegraphics[width=\columnwidth]{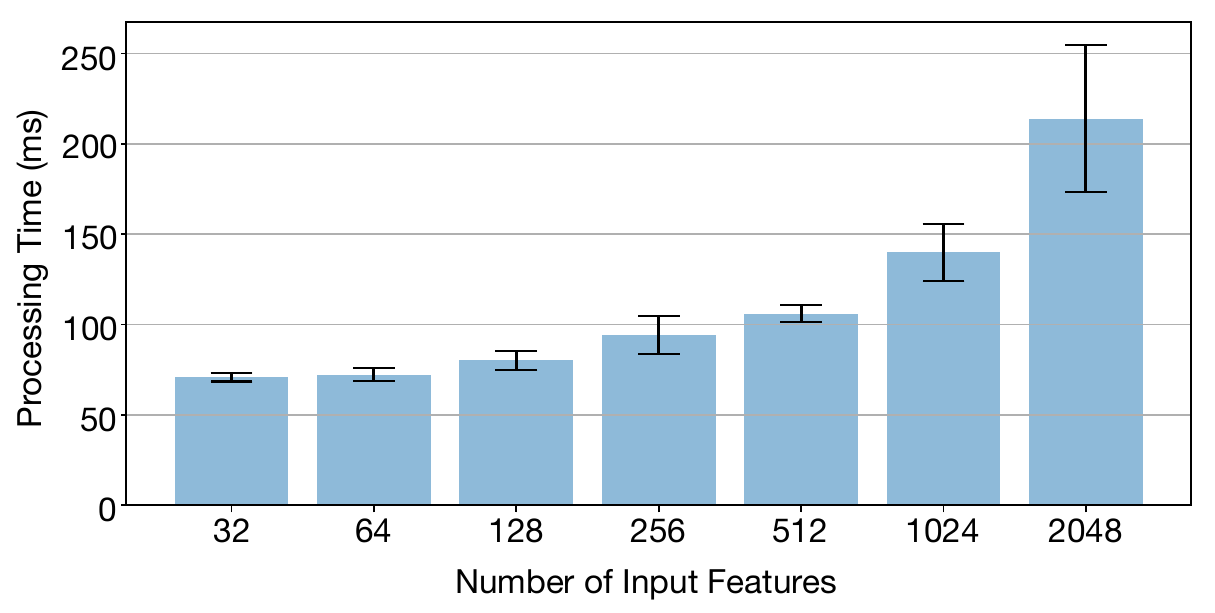}}
		\vskip -0.15in
		\caption{With a $3$ layer fully connected neural network with 4865 parameters built with Keras/Tensorflow, assuming input datapoints have $32$ features, it takes more than $50$ms to obtain a singal output, when using single CPU core (Intel Xeon CPU E5-2690 v3 at 2.60GHz).}
		\label{fig:ml-overhead}
	\end{center}
	\vskip -0.3in
\end{figure}

\textbf{Network LBs deal with high flow arrival rates (higher than $500$ flows/s)--load balancing decisions have to be made within sub-ms or micro-second level~\cite{6lb}}.
Machine learning (ML) and reinforcement learning (RL) approaches are able to make inferences and informed decisions based on multi-modal features extracted from dynamic environments, and they have shown performance gains in various system and networking problems and help avoid error-prone manual configurations~\cite{auto2018sigcomm,decima2018,drl-udn-2019,sivakumar2019mvfst,spotlight2018}.
However, Figure~\ref{fig:ml-overhead} shows that it is computationally intractable to apply ML/RL algorithms on network load balancing problems to make more than $500$ load balancing decisions per second even using just a minimal size of neural network.
Therefore, state-of-the-art network LBs rely on heuristics for decisions on where to place workloads~\cite{maglev,6lb,incab2018}.

This paper formulates the network load balancing problem as a cooperative game to bridge the gap between the networked computing system community and MARL community.
A new load balancing method RLB is proposed, allowing both (i) taking advantage of the learning and inference capacity of RL algorithms given only partial observations, and (ii) making--at micro-second level--data-driven load balancing decisions that bring performance gains.
Experimental evaluations are conducted by running real-world networking traffic to compare and contrast the proposed mechanism with state-of-the-art (SOTA) load balancing algorithms. 

The contribution of this paper are summarised as follows:
(i) This paper formally defines the network load balancing problem as a cooperative game in Dec-POMDP~\cite{oliehoek2016concise} framework and presents a real-world system implementation for realistic performance evaluations.
(ii) This paper proposes a new mechanism that allows benefiting from MARL algorithms and make sub-ms level data-driven load balancing decisions.
(iii) This paper implements $3$ different learning agents--one based on the MARL algorithm QMIX~\cite{rashid2018qmix} and two other based on independent agents using RL algorithm soft actor-critic (SAC)~\cite{haarnoja2018soft} for solving load balancing tasks with multiple LBs, and evaluates their performance on different scenarios, and compares with $5$ SOTA heuristic load balancing algorithms.
(iv) By analysing the experimental results, this paper implies the potential challenges (\textit{e.g.} scalability, synchronisation among distributed agents) for MARL to solve the network load balancing problem, and suggests future work directions.

\section{Related Work}

\textbf{Network Load Balancing:} Network LBs in modern DCs follow the distributed design as in~\cite{maglev}, where multiple LBs randomly assign servers to incoming tasks using Equal-Cost Multi-Path (ECMP).
In case where servers have different processing capacities, servers are assigned with probabilities proportional to their weights using Weighted-Cost Multi-Path (WCMP).
However, as available server processing capacities change with time in DCs~\cite{dragoni2017microservices}, these statically configured weight may not correspond to the actual processing capacities of servers, and they fail to follow and adapt to the dynamic networking environment~\cite{twf2020,spotlight2018}.
As a variant of WCMP, active WCMP (AWCMP) periodically updates server weights using regression algorithm~\cite{spotlight2018} or threshold-based algorithm~\cite{incab2018}, by actively probing resource utilisation information (CPU, memory, or IO usage).
However, AWCMP requires modifications on every servers to manage communication channels that allow collecting observations.
This incurs additional control messages and management overheads, especially in large and scalable DCs.
Local shortest queue (LSQ) counts the number of distributed jobs on each server~\cite{twf2020}, yet it assumes that all servers share the same processing capacity.
The proposed method in this paper requires no modification in the distributed system, yet it is able to passively extract features from network traffic, which help infer server load states and make adaptive and fair load balancing decisions.

\textbf{MARL for Cooperative Games in Networked Computing Systems:} RL has been applied on scheduling problems~\cite{auto2018sigcomm,decima2018,wu2011novel}, which is similar yet different from network load balancing problems.
Agents in scheduling problems know \textit{a priori} the information of workloads--including the expected job durations, job dependencies, \etc--to be distributed before assigning workloads to different processing queues.
However, network LBs have limited observations on only the subset of network flows they distribute--only the number of flows distributed on each server and the elapsed time of on-going network flows.
Comparing with the networking problems, the jobs in scheduling problems also arrive at lower rates (at second level) and have longer completion time.
The inaccurate observation and highly-frequent decision making process make network load balancing a challenging problem to solve.
Previous work~\cite{yao2021reinforced} has explored single-agent RL for load balacing in a network systems, with a close-to SOTA heuristic performance evaluated in simulations.
For the case with multiple LBs working at the same time, this paper applies and studies MARL algorithms on the network load balancing problem in real-world systems to evaluate server load states based only on local observations extracted from network packets, and to dynamically adapt to time-variant environments.
Researches on centralised training, decentralised execution (CTDE) frameworks show performance gain in multi-agent setups~\cite{rashid2018qmix, hao2022api, yang2020qatten}.
\cite{hao2022api} assumes that agents are homogeneous and therefore interchangable, which does not apply in DC networks where load balancers can be deployed on different hardware infrastructure with topologically different distances from servers.
This paper adapts QMIX~\cite{rashid2018qmix}--which achieves similar performance as in~\cite{yang2020qatten}--in an asynchronous mechanism to make highly frequent load balancing decisions, and to learn from partial observations and solve the network load balancing problem as a cooperative game.
RL-based algorithm has also been applied on other load balancing problems~\cite{mai2020multi, houidi2022constrained, mohajer2020mobility, alsuhli2021deep}.
However, the network load balancing problem studied in this paper is different from link load balancing problems studied in~\cite{mai2020multi, houidi2022constrained}, where link utilisation is to be maximised and load balancers have observations on link utilisations.
As discussed in~\cite{survey_l2lb}, in network load balancing problems--more precisely, Layer-$4$ server load balancing problem is studied in this paper--load balancers have no direct observation on server utilisations.
The network load balancing problem studied in this paper is also different from mobility load balancing problem studied in~\cite{mohajer2020mobility, alsuhli2021deep}, where load balancers operate in celular networks instead of DCs and they do not correspond to all resourcese.
The action and optimisation goal (link failure) are entirely different as well.
Furthermore, these works~\cite{mai2020multi, houidi2022constrained, mohajer2020mobility, alsuhli2021deep} conduct evaluations based on simulations while this paper implements and evaluates MARL-based load balancing algorithms in a real-world testbed.

\section{MARL Network Load Balancing}

This section defines the network load balancing problem and formulates it into a cooperative game.

\subsection{Multi-Agent Load Balancing Problem}

Network load balancing can be defined as allocating a Poisson sequence of network flows with different workloads $w\in\mathcal{W}$--whose unit can be, \eg amount of time to process--on a set of $n$ servers, to achieve the maximal utilisation of the computational capacity of all servers.
The workload $w_j(t)$ assigned on the $j$-th server at time $t$ follows an exponential distribution in practical experiments~\cite{facebook-dc-traffic}. 
Multi-agent load balancing problem considers workload distribution through a number of $m$ LBs, which provide high availability and reliability in modern DCs~\cite{maglev}. 

\textbf{Deterministic Case.} The load balancing method for each LB $i\in[m]$ can be a pure strategy $\pi_i\in\Pi_i$: $\mathcal{W}\rightarrow [n]$. Therefore, a deterministic workload assignment function is: $\mathcal{W}\times[m]\rightarrow [n]$.
The ensembled policy for the whole multi-agent load balancing system is thus $\boldsymbol{\pi}=[\pi_1, \dots, \pi_{m}]\in\boldsymbol{\Pi}=\Pi_1\times\dots\times\Pi_m$.
The processing speed for each server is $v_j, j\in[n]$, \ie the amount of workloads that can be processed per unit time.
The remaining workloads on the $j$-th server ($j\in[n]$) during a time interval $t\in[t_0, t_n)$ is thus: 
\begin{align}
    l_j=\frac{\sum_{i\in[m]}\sum_{t\in[t_0, t_n)}w_{i,j}(t)}{v_j},
\end{align}
where $w_{i,j}(t)$ indicates the workload at time $t$ assigned to the $j$-th server via the $i$-th LB. $l_j$ represents the expected time to finish processing all the workloads on the $j$-th server.

\textbf{Stochastic Case.} Since modern DCs have fan-out topology and $m < n$, using deterministic strategy during a time interval will flood $n$ servers under heavy traffic rate (\eg higher than $500$ flows/s), therefore the stochastic load balancing strategies are more often used in practice~\cite{maglev, silkroad2017}.
The stochastic workload assignment function $\alpha$ is defined as: $\mathcal{W}\times[m]\times[n]\rightarrow [0,1]$, representing the probability of the event that the workload is assigned by a specific LB to a specific server. The expected time to finish all workloads on $j$-th server during the time interval $t\in[t_0, t_n)$ is, $\forall j\in[n]$:
\begin{align}
    l_j=\frac{\sum_{i\in[m]}\sum_{t\in[t_0, t_n)}w_{i}(t)\alpha_{i,j}(t)}{v_j}, \sum_{j=1}^n\alpha_{i,j}(t)=1,
    \label{eq:l_j}
\end{align}
$\alpha_{i,j}(t)$ denoting the probability that LB $i$ chooses server $j$ for load $w$ at time $t$.

\textbf{Objective.} The objective for the whole load balancing system can be defined as finding the optimal ensemble policy:
\begin{align}
    \boldsymbol{\pi}^* = \min_{\boldsymbol{\pi}\in\boldsymbol{\Pi}} c(\boldsymbol{l}), \boldsymbol{l}=\{l_j\}, j\in[n]
    \label{eq:obj}
\end{align}
where $c$ is a cost function depending on the expected task finishing time for all servers $j\in[n]$. 
The definition of \textit{makespan} is $c(l_1, \dots, l_j)=\max_{j\in[n]}\{l_j\}$. 
However, in the practical network load balancing problem, LB agents have no observation over the theoretical makespan for the following reasons: 
\begin{enumerate}
    \item operating at the Transport Layer, LB agents are agnostic to application-level information, thus they cannot estimate the remaining workload on each server but counting only the amount of ongoing jobs;
    \item the expected JCT of networking requests follow long-tail distribution~\cite{facebook-dc-traffic}, which makes it difficult to estimate remaining workload only based on the number of ongoing jobs;
    \item in multi-agent setups for the sake of reliability, LB agents only observe partial networking traffic, which makes their counted number of ongoing jobs partial and inaccurate;
    \item the estimation of makespan uses the $\max$ operator, which may produce large variances when facing dynamic traffic.
\end{enumerate}
This paper thus proposes a different cost function--fairness index--which is proved to be equivalent of the makespan as objective. 
\begin{definition}(Fairness) For a vector of task completion time $\boldsymbol{l}=[l_1, \dots, l_n]$ on each server $j\in[n]$, the linear product-based fairness for workload distribution is defined as:
\begin{align}
    F(\boldsymbol{l}) = F([l_1, \dots, l_n]) = \prod_{j \in [n]}\frac{l_j}{\max(\boldsymbol{l})}
\end{align}
\end{definition}

\begin{proposition}
\label{prop:fairness}
Maximising the linear product-based fairness is sufficient for minimising the makespan:
\begin{align}
    \max F(\boldsymbol{l}) \Rightarrow  \min \max(\boldsymbol{l})
\end{align}
\end{proposition}

\begin{proof}
For a vector of task completion time $\boldsymbol{l}=[l_1, \dots, l_n]$ on each server $j\in[n]$, by the definition of fairness, 
\begin{align}
    \max F(\boldsymbol{l}) &= \max \frac{\prod_{j\in[n]} l_j}{\max_{k\prime\in[n]} l_{k^\prime}}
\end{align}
WLOG, let $l_k = \max_{k^\prime\in[n]} l_{k^\prime}$, then,
\begin{align}
     \max F(\boldsymbol{l}) = \max \prod_{j\in[n], j \neq k} l_j
\end{align}
By means inequality,
\begin{align}
    \left(\prod_{j\in[n], j \neq k}l_j\right)^{\frac{1}{n-1}} \leq \frac{\sum_{j\in[n], j \neq k}l_j}{n-1} = \frac{C-l_k}{n-1}, C=\sum_{j\in[n]} l_j.
\end{align}
with the equivalence achieved when $l_i=l_j, \forall i,j\neq k, i,j\in[n]$ holds.
Therefore,
\begin{align}
    \max F(\boldsymbol{l}) &\Rightarrow \max \frac{C-l_k}{n-1} \\
    & \Leftrightarrow \min l_k\\
    &\Leftrightarrow \min \max_{j\in[n]} l_j
\end{align}
The inverse may not hold since $\max \frac{C-l_k}{n-1}$ does not indicate $\max F(\boldsymbol{l})$, so maximising the linear product-based fairness is sufficient but not necessary for minimising the makespan. This finishes the proof.
\end{proof}

To sum up, the network load balancing can be formulated as a constrained optimisation problem:
\begin{align}
    \label{eq:problem_start}
    maximise \prod_{j \in [n]}&\frac{l_j}{\max(\boldsymbol{l})}\\
    subject \, to \quad l_j&=\frac{\sum_{i\in[m]}\sum_{t\in[t_0, t_n)}w_{i}(t)\alpha_{i,j}(t)}{v_j}\label{eq:cons1}\\
                    \sum_{j=1}^n\alpha_{i,j}(t)&=1\label{eq:cons2}\\
                    \sum_{i=1}^{m}w_i &\leq \sum_{j=1}^{n}v_j\label{eq:cons3}\\
                    \alpha_{i,j}&\in[0, 1], w_i, v_j\in(0, +\infty).    
                    \label{eq:problem_end}
\end{align}
The optimisation cost $c$ in Eq.~\eqref{eq:obj} is transformed to be the product-based fairness $F(\boldsymbol{l})$ due to the Proposition~\ref{prop:fairness}. Constraints \eqref{eq:cons1} and \eqref{eq:cons2} are from Eq.~\eqref{eq:l_j} for stochastic network load balancing. Constraint \eqref{eq:cons3} is a necessary condition to have bounded queue backlog (stability).

\subsection{MARL Methods}
\label{sec:marl}
The multi-agent load balancing problem defined in Eq.~\eqref{eq:problem_start}-\eqref{eq:problem_end} can be viewed as a multi-agent cooperative game, where each agent needs to coordinate their behaviour to maximise the common payoff. Specifically, each agent acts independently according to their local observations, the common payoff is improved as long as each agent improves their local policies. 

\textbf{Dec-POMDP:}
MARL for cooperative games can be formulated as decentralised partially observable Markov decision process (Dec-POMDP)~\cite{oliehoek2016concise}, which can be represented as $(\mathcal{I}, \mathcal{S}, \mathcal{A}, R, \mathcal{O}, \mathcal{T}, \gamma)$. $\mathcal{I}$ is the agent set, 
$\mathcal{S}$ is the state set and $\mathcal{A}=\times_i \mathcal{A}_i, i\in\mathcal{I}$ is the joint action set, $\mathcal{O}=\times_i \mathcal{O}_i, i\in\mathcal{I}$ is the joint observation set, and $R$ is the global reward function $R(s,a)$: $\mathcal{S}\times \mathcal{A}\rightarrow \mathbb{R}$ for current state $s\in\mathcal{S}$ and action $\boldsymbol{a}\in\mathcal{A}$. 
The state-transition probability from current state and action to a next state $s^\prime\in\mathcal{S}$ is defined by $\mathcal{T}(s^\prime|s,\boldsymbol{a})$: $\mathcal{S}\times\mathcal{A}\times\mathcal{S}\rightarrow [0,1]$. $\gamma\in(0,1)$ is a reward discount factor. The goal of the RL algorithm is optimising the joint policy $\boldsymbol{\pi}\in\boldsymbol{\Pi} $ to maximise their expected cumulative rewards: $\max_{\boldsymbol{\pi}\in\boldsymbol{\Pi}}\mathbb{E}_{\boldsymbol{\pi}}[\sum_t \gamma^t r_t]$.

To solve the above Dec-POMDP problem, this paper implements and compares $3$ different RL schemes: (i) CTDE, (ii) centralised training and execution (single agent), and (iii) independent agents.

\textbf{QMIX:} QMIX~\cite{rashid2018qmix} algorithm is implemented  in the proposed method, in a CTDE manner. Specifically, QMIX estimates a total $Q$-value function $Q_{tot}$ as a nonlinear combination of the $Q_i$-value for each agent $i\in\mathcal{I}$, as long as the monotonic dependence relationship is satisfied: $\frac{\partial Q_{tot}}{\partial Q_i}, \forall i\in[\mathcal{I}]$. $Q_{tot}(\boldsymbol{\tau}, \boldsymbol{a}, s)$ is a function of joint action-observation history $\boldsymbol{\tau}$, joint action $\boldsymbol{a}$ and the state $s$, while  $Q_i(\tau_i, a_i)$ for each agent is a function of agent observed history $\tau_i$ and its own action $a_i$. The update rule of QMIX follows: \[\min L=\min\sum[Q_{tot}(\boldsymbol{\tau}, \boldsymbol{a}, s)-(r+\gamma \max_{\boldsymbol{a}'} Q_{tot}(\boldsymbol{\tau}', \boldsymbol{a}', s') )]^2.\]
Each LB agent using QMIX algorithm has a stochastic policy. 

\textbf{SAC:} For single-agent game, soft actor-critic (SAC)~\cite{haarnoja2018soft} follows the maximum entropy reinforcement learning framework, which optimises the objective $\mathbb{E}[\sum_t \gamma^t r_t+\alpha \mathcal{H(\pi_\theta)}]$ to encourage the entropy $\mathcal{H}(\cdot)$ of the policy $\pi_\theta$ during the learning process. Specifically, the critic $Q$ network is updated using the gradients $\nabla_\phi\mathbb{E}_{s,a}\bigg[\bigg(Q_\phi(s,a)-R(s,a)-\gamma \mathbb{E}_{s^\prime}[V_{\tilde{\phi}}(s^\prime)]\bigg)^2\bigg]$, where $V_{\tilde{\phi}}(s^\prime)=\mathbb{E}_{a^\prime}[Q_{\tilde{\phi}}(s^\prime, a^\prime)-\alpha\log\pi_\theta(a^\prime|s^\prime)]$ and $Q_{\tilde{\phi}}$ is the target $Q$ network; the actor policy $\pi_\theta$ is updated using $\nabla_\theta\mathbb{E}_s[\mathbb{E}_{a\sim\pi_\theta}[\alpha\log\pi_\theta(a|s)-Q_\phi(s,a)]]$.
Single-agent SAC (S-SAC) method is implemented as the second RL scheme.

\textbf{Independent Learning.}
Apart from QMIX, independent learning agents treat the objective in Eq.~\eqref{eq:obj} from an independent view, where the optimal ensemble policy is factorised as the optimization over each individual policy for each LB agent:
\begin{align}
    \pi_i^* = \min_{\pi_i\in\Pi_i, i\in[m]} c(l_{i}), i\in[m]
    \label{eq:iql_obj}
\end{align}
where $l_i=\{l_{i,j}\}, l_{i,j}=\frac{\sum_{t\in[t_0, t_n)}w_{i}(t)\alpha_{i,j}(t)}{v_j}, j\in[n]$ for the stochastic case. The objective achieved with Eq.~\eqref{eq:iql_obj} is different from Eq.~\eqref{eq:obj} unless the workloads going to each LB are the same at all time, which is impossible in practice. SAC is used for each independent LB agent, which gives the independent-SAC (I-SAC) method.


\begin{figure*}[t]
	\begin{center}
		\centerline{\includegraphics[width=1.9\columnwidth]{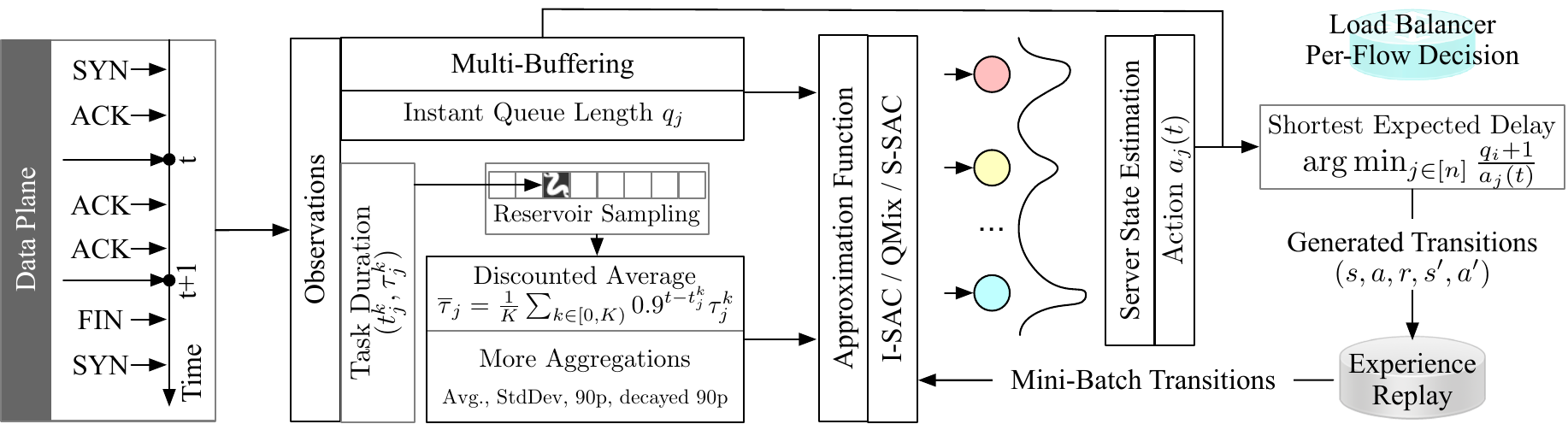}}
		\vskip -0.1in
		\caption{Overview of the proposed MARL framework for network LB: A distributed learning framework with multiple LB agents is implemented to interact with network LB devices and allocate tasks on different servers. Each LB agent contains a replay buffer and can learn using $1$ of the $3$ different RL algorithms--independent SAC (I-SAC), QMix, or single-agent SAC (S-SAC). The $3$ RL algorithms consume the network features (on-going flows and flow duration statistics on each server) as well as the actions from last time step, and they generate server load state estimations as the next time-step action for making fair per-flow-level decision based on the shortest expected delay algorithm.}
		\label{fig:design}
	\end{center}
	\vskip -0.2in
\end{figure*}

\subsection{MARL for Multi-Agent Load Balancing}

\alglanguage{pseudocode}
\begin{algorithm}[htbp]
\caption{MARLLB}
\label{alg:marllb}
\begin{algorithmic}[1]
\State \textbf{Initialise:}
\State \quad replay buffer $\mathcal{B}$
\State \quad learning agents parameterised by $\boldsymbol{\theta}=\{\theta_i\}, \forall i \in [m]$
\State \quad reinforcement learning algorithm $\mathcal{P}$
\State \quad server processing speed function $v_j, \forall j \in [n]$
\State \quad initial observed instant queue length on server $k$ by the $i$-th LB: $q_{i,k}=0,  \forall i\in[m], k\in[n]$
\While {not converge}
\State \textbf{Reset:}
\State \quad server load state $X_j(1) \gets 0, \forall j\in[n]$
\State \quad observations $\boldsymbol{o}(1)$ for LB agents
\For {$t=1,\dots, N$}
    
    \For {LB agent $i$}
        \State $w_{i,j}(t) \gets 0, \forall j\in[n]$
        \State {$a_{i}(t) \gets \{a_{i,j}(t)\}_{j=1}^n =\pi_{\theta_i}(\boldsymbol{o}_{i}(t))$}
    \EndFor
    
    \For {job $\tilde{w}$ arrived at LB $i$ between timestep [$t$, $t+1$)}
        \State LB $i$ assigns $\tilde{w}$ to server $j = \arg \max_{k\in[n]}\frac{q_{i,k}+1}{a_{i,k}(t)}$
        \State {$w_{i,j}(t) \gets w_{i,j}(t) +\tilde{w}$}
    \EndFor
    
    \For {each server $j$}
        \State {$X_j(t+1) \gets max\{X_j(t)+\sum_{i=1}^{m}w_{i,j}(t) - v_j, 0\}$}\Comment{update workload}
    \EndFor
    \State Receive reward $r(t)$
	\State Collect observation $\boldsymbol{o}(t+1)$
    \State $\mathcal{B}=\mathcal{B}\bigcup (\boldsymbol{o}(t), \{a_i(t)\}, r(t), \boldsymbol{o}(t+1))$ \Comment{Update replay buffer}
\EndFor
\State $\boldsymbol{\theta} \gets \mathcal{P}(\mathcal{B})$\Comment{Update agents with RL/MARL algorithms}
\EndWhile
\Return $\boldsymbol{\theta}$
\Statex
\end{algorithmic}
  \vspace{-0.4cm}
\end{algorithm}

The network load balancing problem belongs to multi-commodity flow problems and is NP-hard, which makes it hard to solve with trivial heuristic solution at a micro-second level speed~\cite{sen2013scalable}.
In real-world systems, limited observations on system states and changing environments require LB agents to continuously approximate server load states.
This section describes the network load balancing problem mathematically as a cooperative Dec-POMDP under realistic constraints.
The overview of the MARL framework is depicted in Figure~\ref{fig:design}. The pseudo-code of MARL framework for network load balancing is shown in Algorithm.~\ref{alg:marllb}.

\textbf{Agent Set.} There is a set of homogeneous LB agents $\mathcal{I}$ ($|\mathcal{I}| = m$) distributing workloads among the same set of $n$ application servers. Each agent only distributes and observes over a subset of workloads that arrive at the system. 

\textbf{State and Observation Set.}  The state set is defined as $\mathcal{S} = \mathcal{W} \times \mathcal{V}$, where $\mathcal{W} = {\boldsymbol{w}: \boldsymbol{w} \in (0, \infty)^{m}}$ is a set of incoming network traffic (workloads) to be distributed among servers, and $\mathcal{V} = {\boldsymbol{v}: \boldsymbol{v} \in (0, \infty)^{n}}$ is a set of server processing speeds.
The observation set $\mathcal{O} = {(\boldsymbol{q}, \boldsymbol{\tau}): \boldsymbol{q}, \boldsymbol{\tau} \in (0, \infty)^{n}}$, where $\boldsymbol{q}$ is a vector of counting numbers, each represents the number of on-going task on each server, and $\boldsymbol{\tau}$ is a vector of statistical evaluations (mean, standard deviation, $90$th-percentile, and discounted mean and $90$th-percentile over time) of task elapsed time on each server.

\textbf{Action Set.} $\mathcal{A}=\times_i \mathcal{A}_i$ is the action set containing the individual action set $\mathcal{A}_i$ for each agent $i\in\mathcal{I}$. In the discrete action set $\mathcal{A}_i\subset \mathbb{R_+}^n$, an action $a_i$ is a vector representing the weights for the current workload to be allocated to $n$ application servers by LB agent $i$.
To make hundreds or thousands of load balancing decisions per second while incorporating RL intelligence, this paper adopts the form of the shortest expected delay (SED)\footnote{http://www.linuxvirtualserver.org} to assign server $\arg \min_{j\in[n]}\frac{{q}_{i,j}+1}{a_{i,j}(t)}$ to the newly arrived flow, where $q_{i,j}$ is the number of on-going flows on server $j$ observed by LB $i$, and $a_{i,j}(t)$ is the weight assigned to server $j$ by LB $i$ at timestep $t$.

Conventionally, LB agents make actions on receipt of each networking requests, assigning a server to new-coming requests.
However, it is not possible for RL models to make such decisions under extremely high traffic rates in practice--sub-ms interval between two consecutive decisions.
Using the form of the SED, this paper converts the action of LB agents from assigning servers for each request to periodically (250ms)  and dynamically estimating server processing speed $a_{i,j}(t)$.
By tracking the number of on-going tasks on receipt of every network flow, this allows the proposed LB agents to make load balancing decisions at the pace of task arrival rates and guarantees high-throughput, while adapting to the ever-changing and dynamic server load states and networking environments.

\textbf{State and Observation Transition.}
The state transition probability function is defined as $\mathcal{T}: \mathcal{S}\times\mathcal{A}\times\mathcal{S}\rightarrow [0,1]$, following Markov decision process. 
More specifically, $\mathcal{T}(\boldsymbol{s}_{t+1}|\boldsymbol{s}_t, \boldsymbol{a}_t) = Pr(\boldsymbol{s}_{t+1}|\rho(\boldsymbol{s}_{t}, \boldsymbol{a}_t))$, where $\boldsymbol{a}_t\in\mathcal{A}$, $\boldsymbol{s}_t, \boldsymbol{s}_{t+1}\in\mathcal{S}$,
$\rho(\boldsymbol{s}_{t}, \boldsymbol{a}_t) \mapsto \boldsymbol{s}_{t+\delta t}$ represents the response of servers given the updated workloads distribution, and $Pr(\boldsymbol{s}_{t+1}|\boldsymbol{s}_{t+\delta t})$ represents the change of incoming traffic rates and server processing speeds.
The time interval between two actions is denoted as $\Delta t = 250$ms and $\delta t \ll \Delta t$.

\alglanguage{pseudocode}
\begin{algorithm}[tbp]
\caption{Reservoir sampling}
\label{alg:feature-reservoir}
\begin{algorithmic}[1]
\State $K \gets$ reservoir buffer size
\State $p \gets$ probability of gathering samples
\State $buf \gets [(0, 0), \dots, (0, 0)]$\Comment{Size of $K$}
\State $M \gets \frac{1}{p}$
\For {each observed sample $v$ arriving at $t$}
	\State $randomId \gets rand()$
	\If {$randomId\%M == 0$}
	    \State $idx \gets randomId\%N$ \Comment{randomly select one index}
	    \State $buf[idx] \gets (t, v)$ \Comment{register sample in buffer}
	\EndIf
\EndFor
\Statex
\end{algorithmic}
  \vspace{-0.4cm}%
\end{algorithm}

\begin{figure}[tbp]
	\begin{center}
		\vskip -0.18in
		\centerline{\includegraphics[width=.95\columnwidth]{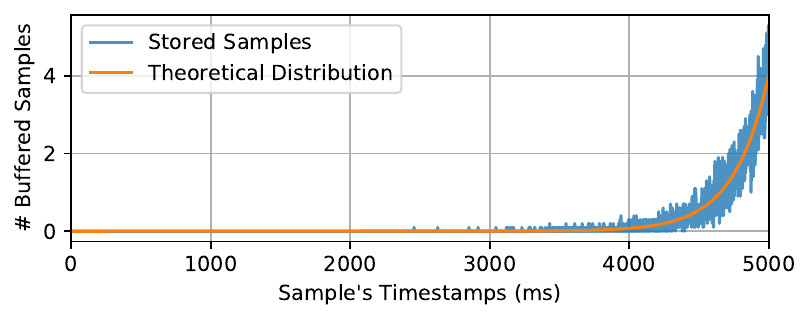}}
		\vskip -0.15in
		\caption{An example of reservoir samples' timestamp distribution with $\lambda=80$, $p = 0.05$, $K=10000$, $n\leq50000$.}
		\label{fig:feature-reservoir}
	\end{center}
	\vskip -0.35in
\end{figure}

\textbf{Observation Probability Function and Feature Collection.}
The observation probability function $\Omega: \mathcal{S}\times\mathcal{A}\times\mathcal{O}\rightarrow[0, 1]$ describes the measurement errors that can occur when extracting and collecting features and statistics from network packets on LB agents.
The counters of on-going network flows $\boldsymbol{q}$ are tracked based on connection states (\eg identified by TCP SYN, FIN packets).
These counters are subject to partial observations in presence of multiple LB agents.
To reduce the impact of partial observation, this paper proposes to use the flow duration (elapsed time since the connection establishment) to indicate the server load state.
The intuition in behind is that, for the same service provided by the server cluster, heavy-loaded or less powerful servers yield longer flow duration than less-loaded or more powerful servers.
Using reservoir sampling (Algorithm~\ref{alg:feature-reservoir}), an exponentially-distributed number of flow duration samples are collected over time.
For a Poisson stream of events with an arrival rate $\lambda$, the expectation of the amount of samples that are preserved in buffer after $T$ steps is $E = \lambda p \left(\frac{K - p}{K}\right)^{\lambda T}$, where $p$ is the probability of gathering sample and $k$ is the size of reservoir buffer.
An example reservoir samples distribution over time is shown in Figure~\ref{fig:feature-reservoir}.
Flow duration samples can be gainfully used to infer server load state and reduce the impact of partial observations in presence of multiple LB agents.

\textbf{Reward Function.}
Since LB agents have limited observations over the actual server load states $\boldsymbol{s}$, the paper uses the flow duration to approximate the sum of queuing delay and task workload over the underlying processing speed of a server.
To give more credits to the latest observations, given the set of samples $\{(t_j^k, \tau_{j}^{k})|k \in [K]\}$ of server $j$, the discounted average of flow elapsed time (an estimatiion of $l_j$ in Eq.~\eqref{eq:l_j}) at time $t$ is computed as $\overline{\tau}_{j}(t) = \frac{1}{K}\sum_{k \in [0, K)}\gamma^{t- t_{j}^{k}}\tau_{i}^{k}$, where $\gamma = 0.9$ in this paper.
Then, based on the Proposition~\ref{prop:fairness}, the reward function is defined as the fairness index of the exponentially weighted average of $\overline{\boldsymbol{\tau}}$ for all servers:
\begin{equation}
  r_{t+1} =
    \begin{cases}
      F(\overline{\boldsymbol{\tau}}_{t}) & \text{if $t = 0$}\\
      F((1-\gamma)\overline{\boldsymbol{\tau}}_{t} + \gamma \overline{\boldsymbol{\tau}}_{t+1})& \text{otherwise.}
    \end{cases}       
\end{equation}

\textbf{Objective Function.} The objective is to maximise their expected cumulative rewards: $\max_{\boldsymbol{\pi}\in\boldsymbol{\Pi}}\mathbb{E}_{\boldsymbol{\pi}}[\sum_t \gamma^t r_t]$, through optimising over the parameterised joint policy $\boldsymbol{\pi}\in\boldsymbol{\Pi}, \boldsymbol{\pi}=\times_{i\in\mathcal{I}}\pi_i,\pi_i:\tilde{\mathcal{O}}_i\times\mathcal{A}_i\rightarrow[0,1]$ is the stochastic policy for agent $i$. $\tilde{\mathcal{O}}_i$ is a concatenation of historical observations and actions for agent $i$. This paper uses gated recurrent units (GRU)~\cite{chung2014empirical} for both QMIX and SAC agents to handle the sequential history information. 

\section{Implementation}

\begin{figure}[tbp]
	\begin{center}
		\vskip -0.1in
		\centerline{\includegraphics[width=\columnwidth]{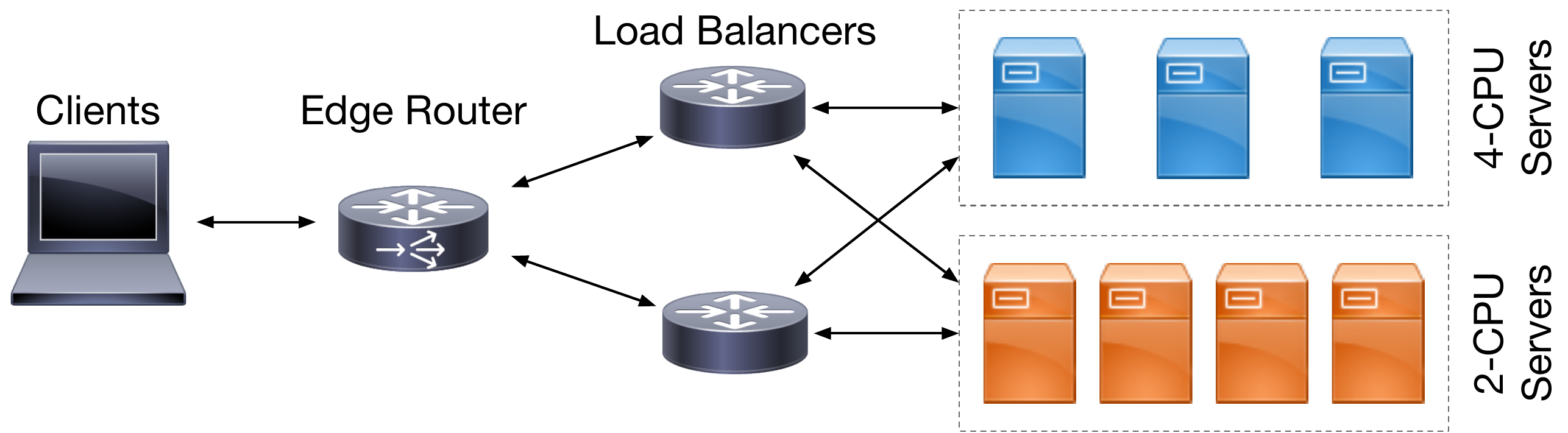}}
		\vskip -0.1in
		\caption{The moderate-scale testbed topology consisting of $1$ traffic generator representing clients, an edge router, $2$ LBs and $7$ application servers with different processing capacities.}
		\label{fig:exp-testbed}
	\end{center}
	\vskip -0.3in
\end{figure}

To evaluate the performance of MARL LB algorithms in different realistic setups, experiments are conducted in a real-world system with real network traces deployed on physical servers.
The experimental platform consists of client nodes, an edge router nodes, LB agents, and Apache HTTP servers providing Web services, virtualised as Kernel-based Virtual Machines (KVMs) as in real-world cloud environments, with the same topology as in Figure~\ref{fig:exp-testbed}.

\subsection{System Platform}

The KVMs are virtualised on $4$ UCS B200 M4 servers, each with one Intel Xeon E5-2690 v3 processor ($12$ physical cores and $48$ logical cores), interconnected by UCS 6332 16UP fabric.
Operating systems are \texttt{Ubuntu 18.04.3 LTS} (\texttt{GNU/Linux 4.15.0-128-generic x86\_64}).
The programmable software network stack \texttt{VPP v20.05} is used to implement the network layer (the data plane) of LB agents for feature collection and policy updates.
The KVMs are deployed on the same layer-$2$ link, with statically configured routing tables.

\subsection{Apache HTTP Servers}

Apache HTTP servers share the same VIP address on one end of GRE tunnels with the load balancer on the other end.
The Apache servers use \texttt{mpm\_prefork} module to boost performance.
Each server has max $32$ worker threads.
The TCP backlog is configured as $128$.
The \texttt{tcp\_abort\_on\_overflow} flag is set, so that, in the Linux kernel, when the TCP connection backlog is full, a TCP RST is sent directly to signify the termination of the connection, instead of silently dropping the packet and waiting for a SYN retransmit.
This configuration allows measuring flow completion time as the application response delays without taking into account additional TCP SYN retransmission delays. 

\subsection{$24$-Hour Wikipedia Replay Trace}

In order to evaluate MARL algorithms using real-world environments, this paper creates replicas of Wikipedia servers using the instance of MediaWiki\footnote{https://www.mediawiki.org/wiki/Download} of version $1.30$.
On each application server instance, a MySQL server and the \texttt{memcached} cache daemon are installed. 
To populate MySQL databases, this paper uses the \textit{WikiLoader} tool and a copy of the English version of Wikipedia database. 
The $24$-hour trace, for privacy reasons, does not contain any information that exposes user identities.

\subsection{PHP \texttt{for}-Loop Trace}

Besides the $24$-hour Wikipedia replay trace which is based on MySQL, a PHP \texttt{for}-loop script is created to study CPU-bound applications.
The number of \texttt{for}-loop iterations \texttt{\#iter} directly determines the expected workload for each request.
The data size that is transmitted for each task is also proportional to the number of iterations, which follows an exponential distribution.
This allows to generate a distribution of flow durations and transmitted bytes that preserve the long-tail characteristic of network traffic~\cite{facebook-dc-traffic}.

\begin{table}[t]
	\centering
	\caption{Two testbed configurations}
	\vskip -.15in
	\resizebox{\columnwidth}{!}{ 
	\begin{tabular}{ccc}
		\toprule
		\multicolumn{1}{c|}{\begin{tabular}[c]{@{}c@{}}Testbed Configuration $n$\end{tabular}} & \multicolumn{1}{c|}{Moderate Scale} & \multicolumn{1}{c}{Large Scale} \\ \midrule
		\multicolumn{1}{c|}{Server Group $1$}                                                   & \multicolumn{1}{c|}{$4 \times 2$-CPU}  & \multicolumn{1}{c}{$12 \times 4$-CPU} \\ \hline
		\multicolumn{1}{c|}{Server Group $2$}                                                   & \multicolumn{1}{c|}{$3 \times 4$-CPU}  & \multicolumn{1}{c}{$12 \times 8$-CPU} \\ \hline
		\multicolumn{1}{c|}{LB Agents}                                                   & \multicolumn{1}{c|}{$2 \times 8$-CPU}  & \multicolumn{1}{c}{$6 \times 8$-CPU}  \\ \hline
		\multicolumn{1}{c|}{Network Trace}                                                   & \multicolumn{1}{c|}{Wikipedia Replay}  & \multicolumn{1}{c}{Poisson Traffic}   \\ \hline
		\multicolumn{1}{c|}{Traffic Rates}                                                & \multicolumn{1}{c|}{$[518.8, 796.3]$}   & \multicolumn{1}{c}{$[391.5, 436.7]$}  \\ \hline
        \multicolumn{1}{c|}{JCT Distribution}                                                & \multicolumn{1}{c|}{Real-world distribution}   & \multicolumn{1}{c}{$\exp(200ms)$}  \\ 		
		\bottomrule
	\end{tabular}
	}
	\label{tab:exp-testbed-conf}
	\vskip -0.2in
\end{table}

\subsection{Network Settings}

Two configurations are implemented to study both moderate- and large-scale DC network environments, which is noted in Table~\ref{tab:exp-testbed-conf}.
In the moderate-scale configuration, network trace samples are extracted and replayed from a real-world $24$-hour replay~\cite{6lb}, which consists of requests for CPU-intensive Wiki pages\footnote{Wiki pages are identifiable by the string \texttt{/wiki/index.php/} in URLs.} 
and IO-intensive static pages.
In the large-scale configuration, a synthesised Poisson traffic of CPU-intensive network flows is applied.
The traffic rates of the two network traces in both configurations are selected to consume $80\%\sim95\%$ average provisioned computational resources.

\subsection{MARL Settings}

To apply the QMIX algorithm, the action space is discretised so that the action set for each LB agent $i$ is $\mathcal{A}_i=\{1.0, 1.2, 1.4, 1.6, 1.8, 2.0\}^n$.
QMIX follows a CTDE manner, with each LB having a $Q$ network for specifying the action choice.
In order to implement the centralised training of the QMIX algorithm, TCP sockets are maintained among LB agents. 
Through these TCP sockets, a master LB agent orchestrates the periodic process (every $250$ms) of making actions and collecting observations for all LB agents to create synchronised trajectories for training.
During each episode, each LB agent collects their locally observed system states and rewards.
At the end of each episode, their collected trajectories are merged on the master LB agent for centralised training.
A global reward--the mean of rewards on all LB agents--is computed for each time step as.

\subsection{Benchmark LB Methods}

In experimental evaluations, the QMIX-based MARL (RLB-QMIX) is evaluated and compared against other methods, namely independent SAC (I-SAC) agents, single-agent SAC (S-SAC), and SOTA heuristic methods including Equal-Cost Multi-Path (ECMP)~\cite{silkroad2017}, Weighted-Cost Multi-Path (WCMP)~\cite{maglev}, active WCMP (AWCMP)~\cite{spotlight2018}, Local Shotest Queue (LSQ)~\cite{twf2020}, and SED.
Among these heuristics, WCMP and SED configure server weights proportional to their provisioned CPU power.
AWCMP relies on TCP channels to periodically probe server resource utilisation information (number of busy Apache threads) to update server load state estimation and recompute server weights.
For I-SAC, each LB node has an independent SAC agent with local observations.
There is no communication among LB agents.
Each LB agent follows the independent learning procedure as introduced in Section~\ref{sec:marl}. 
For S-SAC, a single LB node is deployed to distribute and balance all the workloads across the server cluster.
This single LB agent has global observation on system states and it is trained using a SAC algorithm.
The original SAC algorithm works for continuous action spaces only. 
Modifications are made based on~\cite{christodoulou2019soft} to support discrete action space.

\subsection{Hyperparameters and Training Details}
\label{app:hyperparameter}

The hyperparameters for each learning agent (QMIX, I-SAC, S-SAC) and different experimental setups are provided in Table~\ref{tab:rl_params}.
RL-based load balancing methods are trained in both moderate- and large-scale testbed setups for $72$ episodes.
When replay buffer gathers enough samples (more than $25$ episodes of trajectory samples), the LB agents train and update RL models for $25$ iterations before running the next episode.
QMIX and SAC models use the the same neural network architecture for both Q networks and policy networks--$2$ fully-connected layers, followed by $1$ GRU layer, followed by $2$ other fully-connected layers.
The hidden dimension for all layers is $128$.
The activation function of all fully-connected layers is ReLU.
Given the total provisioned computational resource, the traffic rates of network traces for training are carefully selected so that the RL models can learn from sensitive cases where workloads should be carefully placed to avoid overloaded less powerful servers. 
The traffic rates for large-scale setup is lower than the one for moderate-scale setup (see Table~\ref{tab:exp-testbed-conf}), because the synthesised Poisson traffic has heavier per-job workloads than the real-world Wikipedia Web trace.

\begin{table}[t]
\caption{Hyperparameters in MARL-based LB.}
\vskip -0.15in
\label{tab:rl_params}
\begin{tabular}{cccc}
\toprule
                           & \multirow{1}{*}{Hyperparameter} & \multicolumn{1}{c}{Moderate-Scale} & \multicolumn{1}{c}{Large-Scale} \\ \cline{2-4} 
                          
                           & Learning rate                   & $1\times 10^{-3}$  & $3\times 10^{-4}$\\ \cline{2-4} 
                           & Hidden units                    & $128$  & $512$\\ \cline{2-4}
                           & Batch size                      & $12$                & $12$ \\ \cline{2-4} 
                           & Replay Buffer Size              & $3000$              & $3000$ \\ \cline{2-4} 
                           & Episodes                        & $72$                & $72$   \\ \cline{2-4}
                           & Episode Length                  & $60$s               & $30$s  \\ \cline{2-4} 
                           & Step Interval                   & $0.25$s             & $0.25$s \\ \cline{2-4} 
                           & Update Iterations               & $25$                & $25$ \\ \cline{2-4} 
                           & Target Entropy (SAC)            & $-|\mathcal{A}|$    & $-|\mathcal{A}|$ \\ \midrule
\end{tabular}
\vskip -0.2in
\end{table}

\begin{figure}[t]
	\begin{center}
		\centerline{\includegraphics[width=\columnwidth]{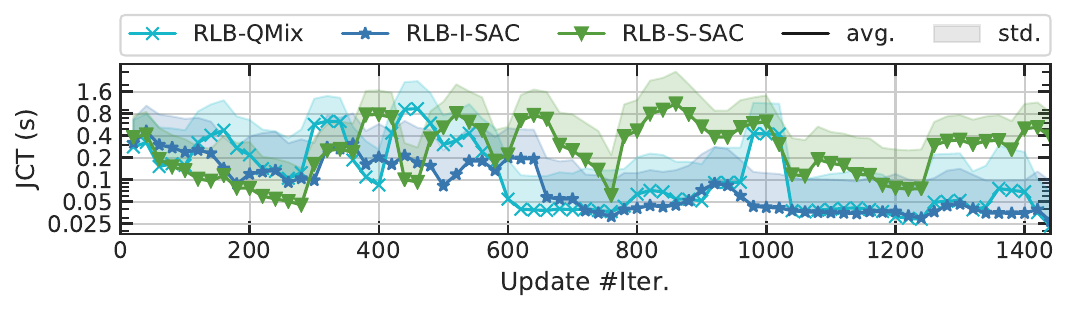}}
		\vskip -0.15in
		\caption{JCT distribution during training using $3$ different RL algorithms.}
		\label{fig:exp-wiki-train}
	\end{center}
	\vskip -0.28in
\end{figure}

\begin{figure}[t]
	\begin{center}
		\centerline{\includegraphics[width=\columnwidth]{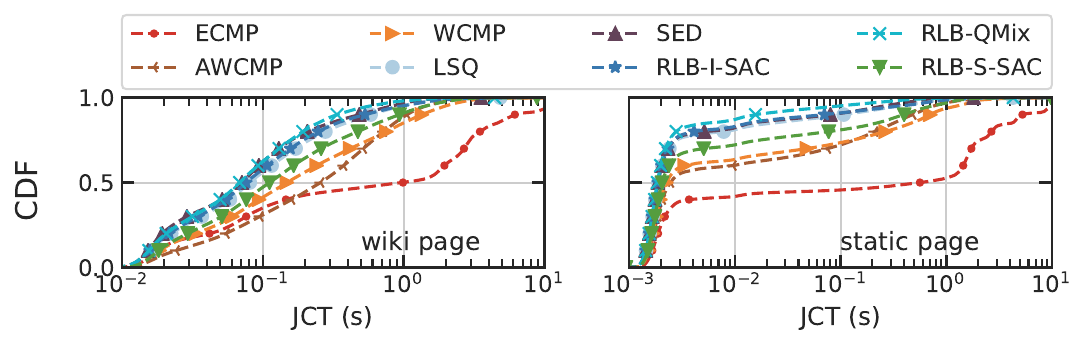}}
		\vskip -0.15in
		\caption{JCT comparison using different load balancing algorithms under different traffic rates (more than $600$ queries/s, $5$ runs per traffic rate).}
		\label{fig:exp-wiki-cdf}
	\end{center}
	\vskip -0.28in
\end{figure}

\section{Evaluation and Results}
\begin{figure}[htbp]
	\begin{center}
		\centerline{\includegraphics[width=\columnwidth]{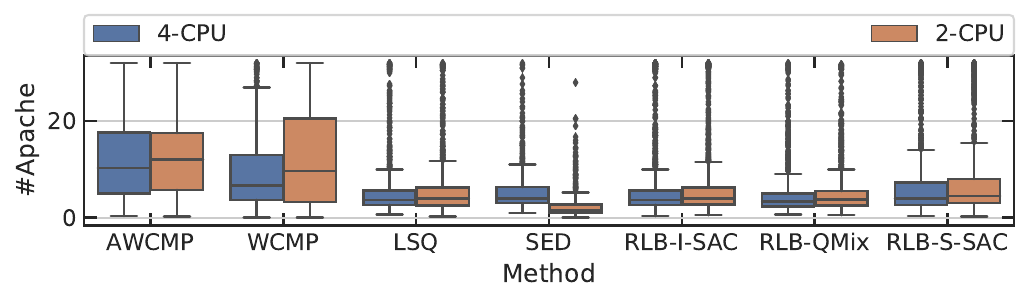}}
		\vskip -0.15in
		\caption{Comparison of number of busy Apache threads on two groups of application servers with different processing capacities.}
		\label{fig:exp-wiki-apache}
	\end{center}
	\vskip -0.25in
\end{figure}

\subsection{Moderate-Scale Testbed Evaluations}

As depicted in Figure~\ref{fig:exp-wiki-train}, MARL-based LB methods show improved performance after $600$ iterations of updates during training while the single agent RLB-S-SAC struggles to learn.
Trained RL-based LB methods are then compared with all the heuristic LB methods on $4$ unseen network traces, which covers a various range of traffic rates from $518.8$ to $796.3$ flows/s.
As shown in in Tables~\ref{tab:compare_small_scale_518} to \ref{tab:compare_small_scale_796}, RLB-QMIX achieves superior performance over all other methods in most scenarios, including the SOTA heuristic method (SED) and other learning agents (I-SAC and S-SAC).
Only when the system is subject to the highest traffic rate (796.3 flows/s), the SED for Wiki pages and the I-SAC agents for static pages win over RLB-QMIX by a slight margin.
Figure~\ref{fig:exp-wiki-cdf} depicts the overall performance comparisons by aggregating the JCTs over the $4$ tested scenarios.
RLB-QMIX is $1.44\times$ and $5.11\times$ faster than SED at $90$th-percentile, which is an important QoS metric.
Figure~\ref{fig:exp-wiki-apache} shows the distribution of the number of busy Apache threads on two groups of servers.
With manually configured server weights, SED assigns $2.329\times$ more workloads on more powerful servers while RLB-QMIX maintains the equivalence between the two groups of servers.

\begin{table}[tbp]
\centering
\caption{Comparison under traffic rate $518.8$ flows/second.}
\vskip -.1in
\begin{tabular}{c|c|c}
\toprule
\multirow{2}{*}{Method} & \multicolumn{2}{c}{Traffic Type} \\
\cline{2-3}
 & \multicolumn{1}{c|}{Wiki} & \multicolumn{1}{c}{Static} \\
\hline
 ECMP & $529.4\pm{110.5}$ & $258.6\pm{94.4}$ \\
 AWCMP &$140.9\pm{4.1}$  & $27.6\pm{4.0}$\\
 WCMP &  $92.6\pm{16.2}$ & $12.7\pm{8.1}$ \\
 LSQ & $78.7\pm{29.9}$ & $8.4\pm{11.0}$ \\
 SED &  $67.2\pm{5.1}$ & $6.7\pm{3.1}$\\
\textbf{RLB-I-SAC} & $84.3\pm{16.9}$ & $9.4\pm{7.4}$\\
\textbf{RLB-QMix} &  $\mathbf{63.4\pm{3.9}}$ & $\mathbf{3.1\pm{0.1}}$  \\
\textbf{RLB-S-SAC} &  $84.2\pm{13.0}$ & $14.4\pm{14.4}$ \\
\bottomrule
\end{tabular}
\label{tab:compare_small_scale_518}
\end{table}

\begin{table}[tbp]
\centering
\caption{Comparison under traffic rate $690.9$ flows/second.}
\vskip -.1in
\begin{tabular}{c|c|c}
\toprule
\multirow{2}{*}{Method} & \multicolumn{2}{c}{Traffic Type} \\
\cline{2-3}
 & \multicolumn{1}{c|}{Wiki} & \multicolumn{1}{c}{Static} \\
\hline
 ECMP &  $3178.9\pm{615.9}$ &  $2835.3\pm{542.8}$ \\
 AWCMP & $430.7\pm{154.5}$ & $153.3\pm{112.5}$\\
 WCMP &  $443.6\pm{268.1}$ & $201.0\pm{188.5}$ \\
 LSQ &  $236.6\pm{164.4}$ & $69.3\pm{105.7}$\\
 SED & $189.3\pm{118.4}$ & $47.3\pm{48.6}$ \\
\textbf{RLB-I-SAC} &  $349.6\pm{397.8}$ & $152.9\pm{260.5}$\\
\textbf{RLB-QMix} &  $\mathbf{166.9\pm{62.3}}$ & $\mathbf{22.0\pm{14.1}}$ \\
\textbf{RLB-S-SAC} &  $397.6\pm{258.2}$ & $156.3\pm{155.9}$ \\
\bottomrule
\end{tabular}
\label{tab:compare_small_scale_690}
\end{table}

\begin{table}[tbp]
\centering
\caption{Comparison under traffic rate $696.5$ flows/second.}
\vskip -.1in
\begin{tabular}{c|c|c}
\toprule
\multirow{2}{*}{Method} & \multicolumn{2}{c}{Traffic Type} \\
\cline{2-3}
 & \multicolumn{1}{c|}{Wiki} & \multicolumn{1}{c}{Static} \\
\hline
 ECMP &  $2748.5\pm{371.1}$ &  $2424.6\pm{388.3}$   \\
 AWCMP &  $348.3\pm{80.3}$ & $101.3\pm{48.1}$ \\
 WCMP &  $530.7\pm{411.7}$ & $287.1\pm{355.9}$ \\
 LSQ &  $207.9\pm{67.6}$ & $40.3\pm{41.0}$\\
 SED &  $182.6\pm{85.7}$ & $40.0\pm{35.1}$ \\
\textbf{RLB-I-SAC} & $146.4\pm{54.7}$ & $19.8\pm{16.9}$ \\
\textbf{RLB-QMix} &  $\mathbf{88.0\pm{10.4}}$ & $\mathbf{4.0\pm{0.7}}$\\
\textbf{RLB-S-SAC} &  $169.1\pm{56.4}$ & $27.0\pm{24.1}$\\
\bottomrule
\end{tabular}
\label{tab:compare_small_scale_696}
\end{table}

\begin{table}[tbp]
\centering
\caption{Comparison under traffic rate $796.3$ flows/second.}
\vskip -.1in
\begin{tabular}{c|c|c}
\toprule
\multirow{2}{*}{Method} & \multicolumn{2}{c}{Traffic Type} \\
\cline{2-3}
 & \multicolumn{1}{c|}{Wiki} & \multicolumn{1}{c}{Static} \\
\hline
 ECMP & $3018.5\pm{837.3}$ &  $2636.8\pm{859.7}$\\
 AWCMP & $539.1\pm{152.4}$ & $203.6\pm{103.2}$\\
 WCMP & $466.8\pm{269.4}$ & $192.5\pm{181.5}$\\
 LSQ & $208.8\pm{117.5}$ & $50.8\pm{38.0}$ \\
 SED & $\mathbf{150.9\pm{69.2}}$ & $22.8\pm{18.5}$ \\
\textbf{RLB-I-SAC} & $155.0\pm{97.0}$ & $\mathbf{17.5\pm{21.9}}$ \\
\textbf{RLB-QMix} & $188.8\pm{104.7}$ & $38.2\pm{32.1}$ \\
\textbf{RLB-S-SAC} & $398.9\pm{367.3}$ & $163.4\pm{212.3}$ \\
\bottomrule
\end{tabular}
\label{tab:compare_small_scale_796}
\end{table}

\subsection{Large-Scale Testbed Evaluations.}

As shown in Table~\ref{tab:compare_large_scale}, although the best performances are achieved with LSQ for $398.5$ flows/s traffic rate and SED for $419.3$ flows/s traffic rate, MARL methods (QMIX and I-SAC) both have a very close performance to the superior method, which demonstrates a certain level of scalability for these learning-based methods to work in real-world large-scale systems.

\begin{table}[tbp]
\centering
\caption{Comparison under different traffic rates ($398.5, 419.3$ flows/second) with synthesised CPU-intensive Poisson traffic for large-scale system setup.}
\vskip -.1in
\begin{tabular}{c|c|c}
\toprule
\multirow{2}{*}{Method} & \multicolumn{2}{c}{Traffic Rate (flows/second)} \\
\cline{2-3}
 & \multicolumn{1}{c|}{398.5} & \multicolumn{1}{c}{419.3} \\
\hline
 ECMP & $5907.2\pm{550.1}$ & $7841.1\pm{484.7}$  \\
 AWCMP & $467.7\pm{5.5}$ & $595.4\pm{6.7}$\\
 WCMP & $629.9\pm{25.0}$ & $1027.5\pm{65.5}$\\
 LSQ & $\mathbf{332.7\pm{1.6}}$ & $420.2\pm{2.0}$  \\
 SED & $338.6\pm{0.7}$ & $\mathbf{410.3\pm{2.3}}$ \\
\textbf{RLB-I-SAC} & $344.8\pm{2.0}$ &  ${425.3}\pm{1.8}$  \\
\textbf{RLB-QMix} & $340.7\pm{1.5}$ & ${419.7}\pm{2.8}$  \\
\textbf{RLB-S-SAC} & $353.0\pm{3.5}$ & $454.6\pm{8.4}$\\
\bottomrule
\end{tabular}
\vskip -.05in
\label{tab:compare_large_scale}
\end{table}

\begin{figure}[t]
	\begin{center}
		\vskip -0.1in
		\centerline{\includegraphics[width=\columnwidth]{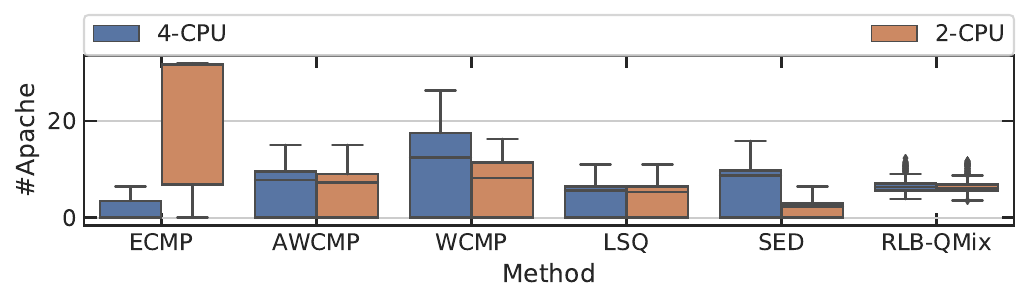}}
		\vskip -0.15in
		\caption{Comparison of the distribution of busy Apache threads on two groups of servers with different processing speeds in the large-scale scenario.}
		\label{fig:exp-add-large-apache}
	\end{center}
	\vskip -0.2in
\end{figure}

Among all the heuristic LB methods, SED has the best performance since it takes both the queue occupation and server processing capacity information into account.
However, when there are multiple LB agents, SED will be mis-guided because of the partially observed numbers of on-going flows.
In the large-scale setup, as depicted in Figure~\ref{fig:exp-add-large-apache}, SED assigns $2.67\times$ and $2.25\times$ more workloads to more powerful servers under $398.5$ and $419.3$ flows/s traffic rates, while the capacity ratio between the two groups of servers is $2$.
This behavior will lead to overloaded powerful servers, which is the reason why LSQ performs better than SED with low traffic rates.
Future studies need to be conducted to learn and adapt to use different strategies under different scenarios.

\begin{figure}[tbp]
	\begin{center}
		\vskip -0.1in
		\centerline{\includegraphics[width=\columnwidth]{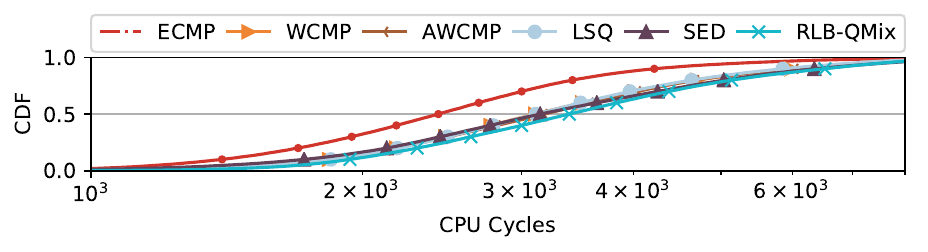}}
		\vskip -0.15in
		\caption{Load balancing decision making latency for each network flow measured under $800$ flows/s traffic rate.}
		\label{fig:exp-latency}
	\end{center}
	\vskip -0.275in
\end{figure}

\subsection{Decision Making Latency}

The decision making latency is compared among all load balancing methods by computing CPU cycles required on the LB node for dispatching every single network flow in the data plane.
As depicted in Figure~\ref{fig:exp-latency}, the RLB-QMIX method has $3.6\%$ and $8.6\%$ additional processing latency than SED and LSQ respectively.
The average number of CPU cycles required for each flow is $4326.27$, which consumes $1.66 \mu s$ on $2.6$GHz-CPU devices.
This allows handling high-throughput network traffic (more than $600$M packet per second) for real-world systems in production.
Therefore, the proposed RL framework for network load balancing problem is able to incorporate intelligence while making high-frequent decisions.


\begin{figure}[t]
	\begin{center}
		\vskip -0.1in
		\centerline{\includegraphics[width=\columnwidth]{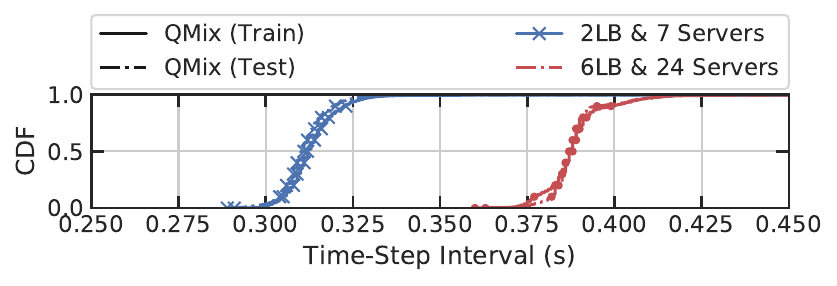}}
		\vskip -0.15in
		\caption{Time-step interval is incremented when using QMIX because of the synchronisation process.}
		\label{fig:exp-add-communication}
	\end{center}
	\vskip -0.3in
\end{figure}

\subsection{Centralised Training and Communication Overhead.}
QMIX adopts the centralised training scheme, which is challenging to implement in real-world distributed system.
This paper synchronises all the LB agents by way of TCP connections among all agents.
Only one master agent takes the responsibility of orchestrating the actions of the other agents so that the interactions between agents and the environments are synchronised and the gathered trajectories follows the Dec-POMDP specification.
This implementation shows good performance in this paper, especially in the moderate-scale setup.
However, in the large-scale setup, RLB-QMIX is outperformed by SED and LSQ with a small margin.
One of the reasons is that the increased communication overhead (latency) and delayed actions at the presence of more LB agents.
As depicted in Figure~\ref{fig:exp-add-communication}, in the large-scale setup, the time interval between two consecutive controls (actions) is $1.24\times$ larger than in the moderate-scale setup.
This additional communication delay fails to effectuate the latest action in time, which deteriorate performance especially in dynamic environments.
Future studies need to be conducted to alleviate this issue.








\section{Conclusions and Future Work}
This paper presents a MARL framework for network load balancing problem, and evaluates different methods for the cooperative game in a real-world system.
The learning-based methods including QMIX, independent-SAC and single-agent SAC are tailored for this application and compared with SOTA heuristic methods.
Experiments show that in moderate-scale system with different traffic rates and types, the MARL method RLB-QMIX achieve superior performance in most settings.
While for large-scale system, learning agents like RLB-QMIX and I-SAC also achieve close performance to the best heuristic methods.
This verify the scalability of the proposed MARL methods for real-world large-scale load balancing system.
Although promising results are achieved, limitations exist in current work: (1) the QMIX algorithm makes additional structural assumption that the joint-action value is monotonic in individual agent value, which may be restrictive for the load balancing problem; (2) reducing the communication cost among agents during training and decision making latency is important for application in real-world load balancing; 
(3) there are other types of scoring mechanism other than the linear-product fairness for load balancing system, like maxinising Jain's fairness, which does not suffice as minimising the makespan yet still worths exploring since it has been used to evaluate load balancing performances~\cite{6lb}. 
Future work includes (1) evaluating more different types of MARL algorithms~\cite{yu2021surprising} on the current system as well as a simulation system in diverse and flexible settings, and (2) the contribution of each proposed component (\eg with or without the GRU), to further improve the performance of MARL solutions.


\bibliographystyle{ACM-Reference-Format}
\balance
\bibliography{reference}

\end{document}